\documentclass[twocolumn,trackchanges]{aastex701}
\usepackage{xspace}
\usepackage{amsmath}
\usepackage{footnote}
\usepackage{booktabs}

\newcommand{\src}{FRB~20220912A\xspace}
\usepackage{CJKutf8}

\shorttitle{Local environment of FRB~20220912A}
\shortauthors{Bhusare et al.}

\begin{document}
\begin{CJK*}{UTF8}{gbsn}

\title{Unveiling the Local Environment of FRB~20220912A: Sub-arcsecond 4–26 GHz Radio Continuum Mapping}

\author[0000-0002-5342-163X]{Yash Bhusare}
\email[show]{ybhusare@ncra.tifr.res.in}

\affil{National Centre for Radio Astrophysics, Tata Institute of Fundamental Research, Post Bag 3, Ganeshkhind, Pune - 411007, India}
\author[0000-0002-0862-6062]{Yogesh Maan}
\email{ymaan@ncra.tifr.res.in}
\affil{National Centre for Radio Astrophysics, Tata Institute of Fundamental Research, Post Bag 3, Ganeshkhind, Pune - 
411007, India}

\author[0000-0002-3615-3514]{Mohit Bhardwaj}
\email{mohitb@iitk.ac.in}
\affiliation{Indian Institute of Technology Kanpur: Kanpur, Uttar Pradesh, IN}

\author[0000-0001-5002-0868]{Thomas C. Abbott}
\email{thomas.abbott@mail.mcgill.ca}
\affiliation{Department of Physics, McGill University, 3600 rue University, Montr\'eal, QC H3A 2T8, Canada}
\affiliation{Trottier Space Institute, McGill University, 3550 rue University, Montr\'eal, QC H3A 2A7, Canada}

\author[0000-0002-9363-8606]{Yuxin Dong (董雨欣)} 
\email{yuxin.dong@northwestern.edu}
\affiliation{Center for Interdisciplinary Exploration and Research in Astronomy, Northwestern University, 1800 Sherman Avenue, Evanston, IL 60201, USA }

\author[0000-0002-5794-2360]{Danté M. Hewitt}
\email{d.m.hewitt@uva.nl}
\affiliation{Anton Pannekoek Institute for Astronomy, University of Amsterdam, Science Park 904, 1098 XH Amsterdam, The Netherlands}

\author[0009-0004-4176-0062]{Afrokk Khan}
\email{afrasiyab.khan@mcgill.ca}
\affiliation{Department of Physics, McGill University, 3600 rue University, Montr\'eal, QC H3A 2T8, Canada}
\affiliation{Trottier Space Institute, McGill University, 3550 rue University, Montr\'eal, QC H3A 2A7, Canada}

\correspondingauthor{Yash Bhusare}

\begin{abstract}

The local environments of repeating fast radio bursts (FRBs) provide critical clues to their progenitors. While some active repeaters (e.g., FRB~20121102A, FRB~20190520B) are embedded in compact persistent radio sources (PRS), others appear to reside in cleaner environments. We present a high-resolution, multi-frequency (4$-$26 GHz) continuum study of the hyperactive repeater FRB 20220912A using the Karl G. Jansky Very Large Array (VLA). We report the discovery of a previously unknown radio source distinct from the compact PRSs seen in other FRBs, spatially coincident with the FRB position and offset by $\approx 300$~mas ($\approx 450$~pc) from the host galaxy's center. The absence of continuum emission in archival milliarcsecond-resolution VLBI observations indicates that the source is resolved out, ruling out a hyper-compact ($< 1$~pc) central-engine-powered origin. We constrain the physical diameter of the emitting region between 75~pc and 190~pc. We further demonstrate that the source is characterized by a steep non-thermal spectral index ($\alpha \approx -0.73$) and a remarkably high star-formation rate surface density $\Sigma_{\text{SFR}} \gtrsim 13~M_{\odot}~\text{yr}^{-1}~\text{kpc}^{-2}$. We argue that this emission is best explained as a compact star-forming region within the host galaxy. This association with a site of ongoing star formation provides strong observational support for the hypothesis that young magnetars, formed after the deaths of massive stars, are the progenitors of at least some repeating FRBs.

\end{abstract}


\section{Introduction}

Fast radio bursts (FRBs) are highly energetic, millisecond-duration extragalactic transients \citep{Lorimer_2007}. Despite their discovery over a decade ago, the emission mechanism of these bursts is poorly understood. Young magnetars have emerged as a leading progenitor candidate due to their high energy reservoir, a key requirement for such emission. Progenitor models involving magnetars gained strong support when a bright FRB-like burst was detected from the Galactic magnetar SGR 1935+2154 \citep{2020Natur.587...54C,2020Natur.587...59B}. Studying the local environments of FRBs provides critical constraints on the progenitor models \citep{2018ApJ...868L...4M,2025ApJ...988..276R}; properties like dispersion measure (DM) and rotation measures (RM) reveal the plasma density and magnetic fields around the source, while persistent radio counterparts can indicate an association with pulsar wind nebula or supernova remnant. 

To date, only five FRBs have been associated with compact persistent radio sources (PRSs): FRB 20121102A \citep{Chatterjee_2017}, FRB 20190520B \citep{Niu_2022}, FRB 20201124A \citep{bruni2024nebularoriginpersistentradio}, FRB 20240114A \citep{bhusare_prs,bruni2024discoveryprsassociatedfrb,2025ApJ...992L..35B}, and FRB 20190417A \citep{2026ApJ...996L..16M}. While most of these sources are confirmed to be compact on parsec scales \citep[$d < 23$ pc,][]{2026ApJ...996L..16M}, the current limits on the physical size of the source associated with FRB 20201124A constrain only down to $d < 700$ pc. Such compact PRSs are interpreted as central-engine-powered sources directly tied to the FRB progenitors \citep[but also see][]{2025PASP..137h4202B}, making them distinct from extended, host-related continuum emission.

With the exception of FRB 20201124A, which displays an inverted spectrum peaking at high frequencies, these PRSs generally exhibit flat radio spectra \citep[$\alpha \approx 0$, see, e.g.,][]{2024ApJ...976..199I}. In contrast, a steep spectral index ($\alpha \approx -0.7$) is typically characteristic of non-thermal radiation from star-forming regions \citep{1992ARA&A..30..575C}. Distinguishing between these spectral properties is thus essential for determining whether the persistent emission is a byproduct of the FRB engine or an associated site of host galaxy star formation.

\begin{table*}[t] 
\centering
\label{tab:astrometry}

\begin{tabular}{@{}lccccc@{}}
\toprule
\textbf{Frequency} & \textbf{Calibrator Offset (RA/Dec)} & \textbf{Source Pos. Error} & \textbf{Beam Size (Minor)} & \textbf{$\Delta$RA } & \textbf{$\Delta$Dec } \\ 
\textbf{(GHz)} & \textbf{(mas)} & \textbf{(mas)} & \textbf{(mas)} & \textbf{(mas)} & \textbf{(mas)} \\ 
\textbf{21.63 (K)}  & +27.05 / -14.56 & 9.10 / 10.71  & 90.0  & -4.76  & -0.62  \\
\textbf{14.71 (Ku)} & +26.99 / -14.62 & 12.11 / 13.11 & 120.0 & -35.26 & -16.87 \\
\textbf{9.92 (X)}   & +26.53 / -15.47 & 17.07 / 17.61 & 170.0 & +9.58  & -12.76 \\
\textbf{6.22 (C)} \tablenotemark{*}   & +27.80 / -14.48 & 27.17 / 32.28 & 270.0 & -164.20 & -39.48 \\ 
\end{tabular}%

\tablenotetext{*} {In C-band, extended emission may contribute to the flux density, which may have introduced large positional deviations when compared with the FRB localization.}

\caption{Astrometric verification of the observations. All units are in milliarcseconds (mas). The systematic offsets in Right Ascension (RA) and Declination (Dec) were derived by comparing the measured position of the phase calibrator (J2322+509) against its cataloged coordinates. `source Pos. Error' represents the formal fitting errors from \textit{imfit}. The measured calibrator offsets were first applied to the target to yield the final residual differences ($\Delta\text{RA}$ and $\Delta\text{Dec}$) between the detected continuum source and the precision VLBI localization of FRB 20220912A. Self-calibration was explicitly omitted to preserve these astrometric measurements.}
\end{table*}

The FRB host environments exhibit significant diversity \citep{2020ApJ...903..152H, 2021ApJ...917...75M, 2024ApJ...971L..51B, 2025ApJ...993..119G}. While many FRBs are found to originate in the spiral arms of star-forming galaxies, they are not strictly confined to regions of active star formation. For instance, recent localizations have placed some FRBs at significant offsets from star-forming knots or the outskirts of quiescent galaxies \citep{2025ApJ...989L..48C, 2025ApJ...979L..21S}, and FRB 20200120E has been localized to a globular cluster \citep{2021ApJ...910L..18B, 2022Natur.602..585K}, challenging the young magnetar formation through Core-collapse supernovae (CCSNe) channel. Moreover, despite extensive searches, no association between FRBs and historical extragalactic supernovae has been identified to date \citep{2025ApJ...992..127L}.

The spatial distribution of FRBs within their hosts offers a window into their formation history \citep{2025ApJ...993..119G}. Significant offsets from the star-forming knots might have resulted from kicks during the progenitor's birth, likely a supernova, or the progenitor might belong to an older stellar population \citep{2021ApJ...908L..12T}. However, identifying such associations requires resolving emission
on intermediate spatial scales of a few tens of parsecs to $\sim1$ kpc. At the typical host galaxy redshifts (e.g., $z \approx 0.077$ for FRB~20220912A), these scales correspond to an angular resolution of approximately $0.05''$ to $0.7''$. Such regions are often too extended to be recovered by the milliarcsecond resolution of Very long baseline interferometry (VLBI), and they are too compact to be isolated from broader galactic emission by telescopes like the upgraded Giant Metrewave Radio Telescope (uGMRT) or the C and D configurations of the Karl G. Jansky Very Large Array (VLA), which typically provide resolutions $\gtrsim 1.5''$ ($\gtrsim 2.2$ kpc).

The known FRB population is observationally divided into apparent one-off events, which constitute the majority, and a smaller subset of repeating sources \citep{2021ApJS..257...59C,2026ApJS..283...34C}. Historically, one-off FRBs have typically been localized with large uncertainties ranging from arcminutes to degrees \citep[e.g.,][]{2021ApJS..257...59C, 2024ApJ...971L..51B}, encompassing the entire host galaxy and blending the FRB's local environment with broader galactic emission. While interferometric advancements such as Australian Square Kilometre Array Pathfinder (ASKAP) \citep{2019Sci...365..565B} and the Canadian Hydrogen Intensity Mapping Experiment (CHIME) Outriggers \citep{2025ApJ...989L..48C} are now enabling sub-arcsecond to milliarcsecond localizations for single bursts, these capabilities are relatively recent. For sources like FRB~20220912A, which was discovered and localized prior to the routine operation of CHIME/FRB Outriggers, its recurrent nature allowed for milliarcsecond localization through VLBI observations. Such high precision enables us to spatially resolve the FRB from its host galaxy center and investigate the immediate circumburst environment. Thus, resolving the host galaxy's structure to study these local environments makes transitioning to high-resolution (sub-arcsec) radio continuum observations essential.

In this paper, we focus on the hyperactive repeater FRB 20220912A. Using high-resolution VLA A-configuration observations across a wide frequency range (4–26 GHz), we report the detection of a radio source spatially coincident with the FRB position. By resolving the emission on sub-arcsecond scales, we investigate the physical nature of the source to determine its origin. The rest of the paper is structured as follows. Section~2 provides more details about our target, FRB~20220912A. Details of the observations are provided in Section~3, followed by data processing details in Section~4. Results are described in Section~5, and we discuss these results in detail in Section~6, which follows an overall summary in Section~7.

\section{\src{}}

\src{} was first discovered by the CHIME/FRB collaboration in September 2022 \citep{2022ATel15679....1M}. Following the discovery, \src{} entered a phase of hyperactivity. Such sustained, high-rate emission is highly unusual even within the repeating FRB population. However, this rarity provided a distinct observational advantage, allowing extensive follow-up across multiple frequencies using different telescopes, including Five hundred meter Aperture Spherical Telescope (FAST), uGMRT, Nançay Radio Telescope, Robert C. Byrd Green Bank Telescope (GBT), Allen Telescope Array and CHIME \citep{zhang2023fastobservationsfrb20220912a,2022ATel15791....1R,2022ATel15806....1B,2024MNRAS.534.3331K,2025arXiv251221889K,2024ApJ...974..296F, 2024MNRAS.52710425S, 2024ApJ...974..170C, 2026arXiv260409098A}. Initial localization was rapidly provided by the DSA-110 array, which associated the bursts with the host galaxy PSO J2309+4842 at a redshift of $z \approx 0.077$ \citep{Ravi_2023}.

 Subsequent monitoring of the burst properties, including polarization, indicated a constant RM of approximately $0 \text{ rad m}^{-2}$  \citep{zhang2023fastobservationsfrb20220912a, 2024ApJ...974..296F}. This near-zero value suggests that the progenitor resides in a likely non-magneto-ionic environment. PRSs are theoretically expected to be found in high magneto-ionic environments \citep{2022ApJ...928L..16Y}. However, the total observed RM is a sum of contributions from the Milky Way, the host galaxy, and the immediate local environment of the source. For this line of sight, the Galactic contribution is estimated at approximately $-15 \text{ rad m}^{-2}$ \citep{2022ATel15679....1M}. While the net RM is low, the local environment could still be complex; for instance, a Milky Way-like host galaxy can contribute several hundred units of RM  \citep{2009ApJ...702.1230T}, potentially underestimating the local contribution. Given the seemingly quiescent magneto-ionic state of \src{}, a correlation between local RM and the luminosity of compact PRS \citep{2022ApJ...928L..16Y} does not predict a luminous PRS. However, the existence of a compact radio counterpart can not really be ruled out. Therefore, imaging remains essential to characterize the immediate surroundings of the source.

Previous radio observations of the field including VLA at C-band (D-configuration) and uGMRT campaigns at 650 MHz and L-band, successfully detected radio emission at the FRB location \citep{bhusare_prs,2024A&A...690A.219P}. However, these detections were limited by large synthesized beam sizes ($> 1.5''$), which prevented the spatial separation of any compact FRB-associated source from the broader emission of the host galaxy. Based on these flux density measurements, \citet{bhusare_prs} derived a spectral index of $\alpha \approx -0.72$, a value consistent with synchrotron emission from a star-forming region.

To refine the localization and search for a compact PRS, \citet{hewitt2023milliarcsecondlocalisationhyperactiverepeating} conducted high-resolution VLBI observations. These observations significantly improved the astrometric accuracy of the burst location ($\sim 5$~mas) but yielded a non-detection of any compact component in the image domain. They also placed a $5\sigma$ upper limit of $80~\mu\text{Jy/beam}$ on any compact radio emission at L-band, effectively constraining the luminosity, L, to be $\leq 1.2 \times 10^{28} \ \mathrm{erg \ s^{-1} \ Hz^{-1}} \ \ (5\sigma)$ of any PRS associated with \src{}.

\begin{table}[ht!]
    \centering
    \caption{Sub-banded Peak Flux Density Measurements}
    \label{tab:flux_measurements}
    \begin{tabular}{lcccc}
        \toprule
        \textbf{Frequency} & \textbf{Peak} & \textbf{Measured} & \textbf{Total} \\ 
        \textbf{} & \textbf{Flux Density} & \textbf{Error (rms)} & \textbf{Error} \\
        \textbf{(GHz)} & \textbf{($\mu$Jy/beam)} & \textbf{($\mu$Jy/beam)} & \textbf{($\mu$Jy/beam)} \\ \midrule
        4.48  & 74.30 & 6.9  & 10.1 \\
        5.51  & 61.80 & 5.1  & 8.1 \\
        6.99  & 51.10 & 4.3  & 6.7 \\
        7.25  & 41.50 & 4.2  & 5.9 \\
        8.48  & 50.00 & 4.8  & 6.9 \\
        9.51  & 44.80 & 3.9  & 5.9 \\
        10.48 & 28.90 & 3.8  & 4.8 \\
        11.51 & 24.70 & 4.5  & 5.1 \\
        12.99 & 38.95 & 4.4  & 5.9 \\
        14.99 & 34.09 & 4.5  & 5.6 \\
        16.99 & $<$34.25\tablenotemark{a} & 6.8  & -- \\
        19.99 & 52.75\tablenotemark{b} & 8.1  & 9.7 \\
        23.99 & $<$47.5\tablenotemark{a} & 9.5  & -- \\ \bottomrule
    \end{tabular}
    \vspace{1mm}
    \begin{minipage}{\columnwidth}
    \tablenotetext{a}{Non-detection; value represents the $5\sigma$ upper limit.}
    \tablenotetext{b}{Measurement affected by a calibration artifact near the 22.2 GHz water line; see Appendix \ref{appendix:calibration_artifact}.}
    \end{minipage}
\end{table}

\section{Observation}

To probe the existence of a faint, compact PRS associated with \src{}, we conducted high-frequency ($4-26$ GHz) observations using the VLA. This high-frequency strategy was motivated by the recent discovery of a PRS associated with FRB~20201124A by \citet{bruni2024nebularoriginpersistentradio}, which exhibits an inverted radio spectrum peaking at higher frequencies.

The initial observation campaign covered the Ku (12--18~GHz) and K (18--26~GHz) bands, and it was conducted during the array's transition from B to A configuration (Project Code: 24B-477) giving resolution of $\approx0.2''$ to $\approx0.1''$. Under this program, we observed in the K-band (18--26~GHz) on 2024 October 12 for 1.15~hours and in the Ku-band (12--18~GHz) on 2024 October 14 for 1.12~hours of on-source time. These observations resulted in the detection of a faint radio counterpart at the position of \src{}.

To robustly constrain the spectral index and determine the physical origin of the emission, we conducted further observations (Project Code: 24B-525) in 2025 February. During this phase, we observed in the X-band (8--12~GHz) on 2025 February 1 for 2.21~hours and in the C-band (4--8~GHz) on 2025 February 3 for 1.88~hours of on-source time. These observations utilized the VLA in its most extended A-configuration to maximize the spatial resolution ($\approx0.4''$ to $\approx0.3''$).

All data were recorded with a spectral resolution of 2.00~MHz. 3C48 served as the primary flux density calibrator, and J2322+509 was utilized as the complex gain (phase) calibrator for all the observations.

\begin{figure}[ht!]
    \centering
    \includegraphics[width=0.48\textwidth]{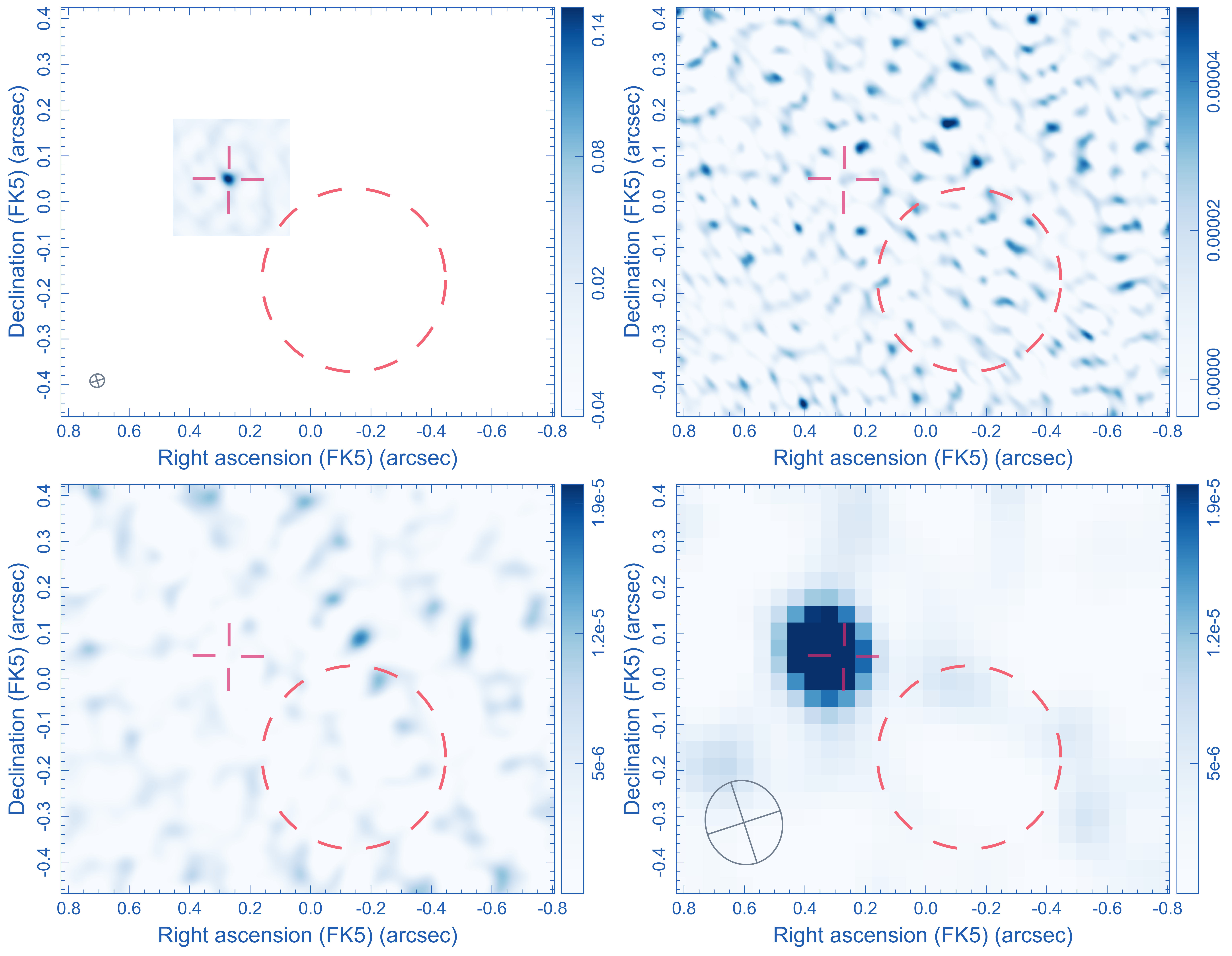} 
    \caption{\textbf{Position Match Analysis.} \textit{Top-Left:} VLBI burst position \citep{hewitt2023milliarcsecondlocalisationhyperactiverepeating}. The field of view (FOV) in this specific panel is highly restricted, meaning the host galaxy nucleus lies entirely outside the insert VLBI image. \textit{Top-Right:} VLBI continuum image showing a non-detection \citep{hewitt2023milliarcsecondlocalisationhyperactiverepeating}, which covers a wider FOV. \textit{Bottom-Left:} VLBI continuum convolved to a 50~mas beam (RMS $6~\mu\text{Jy}$; no compact emission $>30~\mu\text{Jy}$ at $5\sigma$). \textit{Bottom-Right:} VLA X-band full bandwidth image (this work). The red circle marks the host galaxy center (uncertainty ellipse with $0.2''$ semi-major axis). The spatial X-Y coordinates are matched across the images to display the same field. Hollow plus denotes the position of the FRB as measured by \cite{hewitt2023milliarcsecondlocalisationhyperactiverepeating} }
    \label{fig:pos_match}
\end{figure}

\begin{figure*}[ht!]
   \centering
   \includegraphics[width=1.0\textwidth]{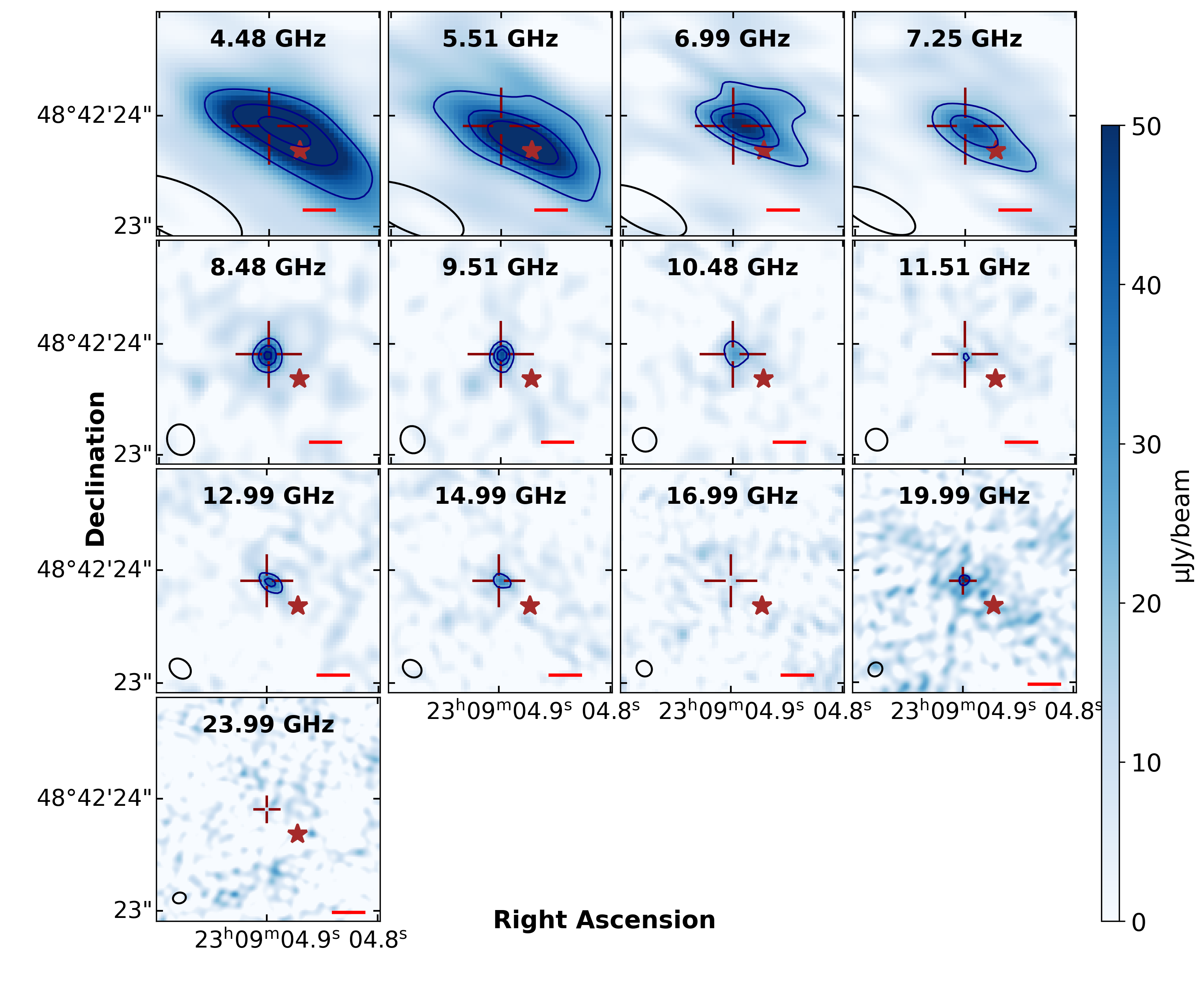} 
   \caption{Multi-frequency VLA images of the persistent source associated with FRB 20220912A. The source is detected in C, X, and lower Ku bands, but drops below the detection threshold at 16.9 GHz and 23.9 GHz. All panels share the same field of view and color scale. Sub-bands were imaged using natural weighting to maximize SNR. The contours on each image is drawn at 5$\sigma$, 8$\sigma$ and 10$\sigma$. Brown star on the image shows host galaxy center and cross sign shows the position of the FRB. The red line in the bottom left is 300 mas $\approx$ 450~pc long.}
   \label{fig:Image}
\end{figure*}

\section{Data Processing}
\subsection{Imaging}
Data reduction was performed using the Common Astronomy Software Applications (CASA) package. We utilized the standard VLA calibration pipeline for initial flagging and calibration. Following the pipeline, imaging was carried out using the \textit{tclean} task. We opted not to perform self-calibration on any of the images; the field lacked sufficient signal at these observation frequencies to provide adequate signal-to-noise ratio (SNR) for gain corrections. Furthermore, this choice was motivated by the astrometric requirements discussed in the Section \ref{sec:astrometry}.

A radio continuum source was detected at the location of \src{} in all the observed bands. The source appears unresolved (at VLA resolution) across the entire frequency range. Furthermore, the integrated flux density is consistent with the peak flux density in all bands, confirming the compact nature of the emission. The peak to integrated flux density ratios are consistent with unity across all bands. For X, Ku, and K-bands, the ratios lie within $1\sigma$ of unity, while in C-band the ratio deviates at $1.6\sigma$ level (only for the C-band image, we used uniform weighting, as robust weighting, i.e., briggs=0.5, produced a significantly broader beam). To model the spectral nature of the source, we divided the individual bands into sub-bands of equal bandwidth, resulting in four sub-bands for the C-band (4–8 GHz) and X-band (8–12 GHz), three sub-bands for the Ku-band (12–18 GHz), and two sub-bands for the K-band (18–26 GHz).

Each sub-band was imaged individually using \textit{tclean} with the \textit{mtmfs} deconvolver and natural weighting. This choice was motivated by the fact that each sub-band contains significantly less data than the full bandwidth; natural weighting compensates for this reduced data volume by providing the lower noise floor (RMS) at the cost of a slightly larger synthesized beam.

 To adequately sample the primary beam, the full-band images were constructed with a size of around 16K$\times$16K pixels. The cell sizes were appropriately scaled to oversample the synthesized beam for each receiver, set to $0.054^{\prime\prime}$, $0.037^{\prime\prime}$, $0.024^{\prime\prime}$, and $0.014^{\prime\prime}$ for the C, X, Ku, and K bands, respectively. For the sub-band imaging, since no other source was present in the field of view, we restricted the image size to 1K$\times$1K pixels to optimize computational efficiency. The cell sizes for these sub-banded images were set to $0.042^{\prime\prime}$, $0.036^{\prime\prime}$, $0.028^{\prime\prime}$, and $0.014^{\prime\prime}$ for the C, X, Ku, and K bands, respectively.

\subsection{Flux density and the spectrum}\label{sec:flux_density_of_source}

To determine the flux density of the radio source at each sub-band, we extracted the peak pixel value from each sub-banded image (deconvolved image). Because the source remains unresolved across all frequencies, the peak pixel value is consistent with the total flux density. 

The flux density measurements, along with the associated measurement (the local rms noise) and the total uncertainties (quadrature addition of measurement uncertainties and $10\%$ systematic uncertainties) are provided in Table~\ref{tab:flux_measurements}. For spectral analysis involving fitting, total uncertainties was consider. We note a significant flux density anomaly at $\sim$19.9 GHz. To determine whether this feature is intrinsic or an instrumental artifact, we analyze the spectral behavior of the complex gain calibrator (J2322+509) observed during the same session. As discussed in the Appendix \ref{appendix:calibration_artifact} (and Figure~\ref{fig:spectra_phase_cal}), the calibrator exhibits enhancement at the same frequency. This correlation suggests that the 19.9 GHz excess is not intrinsic to the source but is a calibration artifact likely due to atmospheric opacity variations near the 22.2 GHz water vapor line. Such artifacts can arise when the atmospheric optical depth differs significantly between the flux density calibrator and the target field, leading to an imperfect transfer of the flux scale. 

To estimate the spectral index, $\alpha$, we modeled the continuum emission of the source with a standard power law of the form, $S_{\nu} \propto \nu^{\alpha}$, where $S_\nu$ is the flux density at frequency $\nu$. The spectral index was derived by fitting a linear model in log-log space to all measured flux densities, excluding upper limits. This fitting was performed using the Nelder-Mead minimisation algorithm implemented within the Python package \textit{lmfit} \citep{2021zndo....598352N}. During the fitting process, each data point was weighted according to its uncertainty.

\subsection{Astrometry}\label{sec:astrometry}

To verify if the detected emission is indeed spatially associated with \src{}, we performed astrometry analysis. Since the target field lacks secondary sources for a direct cross-match with known catalogs, we utilized the phase calibrator (J2322+509) to quantify the systematic positional errors. The phase calibrator was imaged using the same \textit{tclean} parameters and weighting schemes as \src{}. We did not perform self-calibration on any image, as doing so would introduce independent gain solutions for the target, different from those of the phase calibrator, potentially masking the systematic shifts we aim to measure. By comparing the measured position of the calibrator in our images against its cataloged coordinates, we derived the offsets (summarized in Table \ref{tab:astrometry}). These systematic offsets were then applied to the measured target coordinates to determine the final source position.

We further illustrate this spatial coincidence in Figure \ref{fig:pos_match}, which compares our X band image with the VLBI burst localization reported by \citet{hewitt2023milliarcsecondlocalisationhyperactiverepeating}. The figure demonstrates a spatial separation between the detected continuum radio source and the host galaxy's center (represented by the red circle), while showing alignment between the FRB burst position and the continuum emission.

\begin{figure}[ht!]
    \centering
    \includegraphics[width=0.49\textwidth]{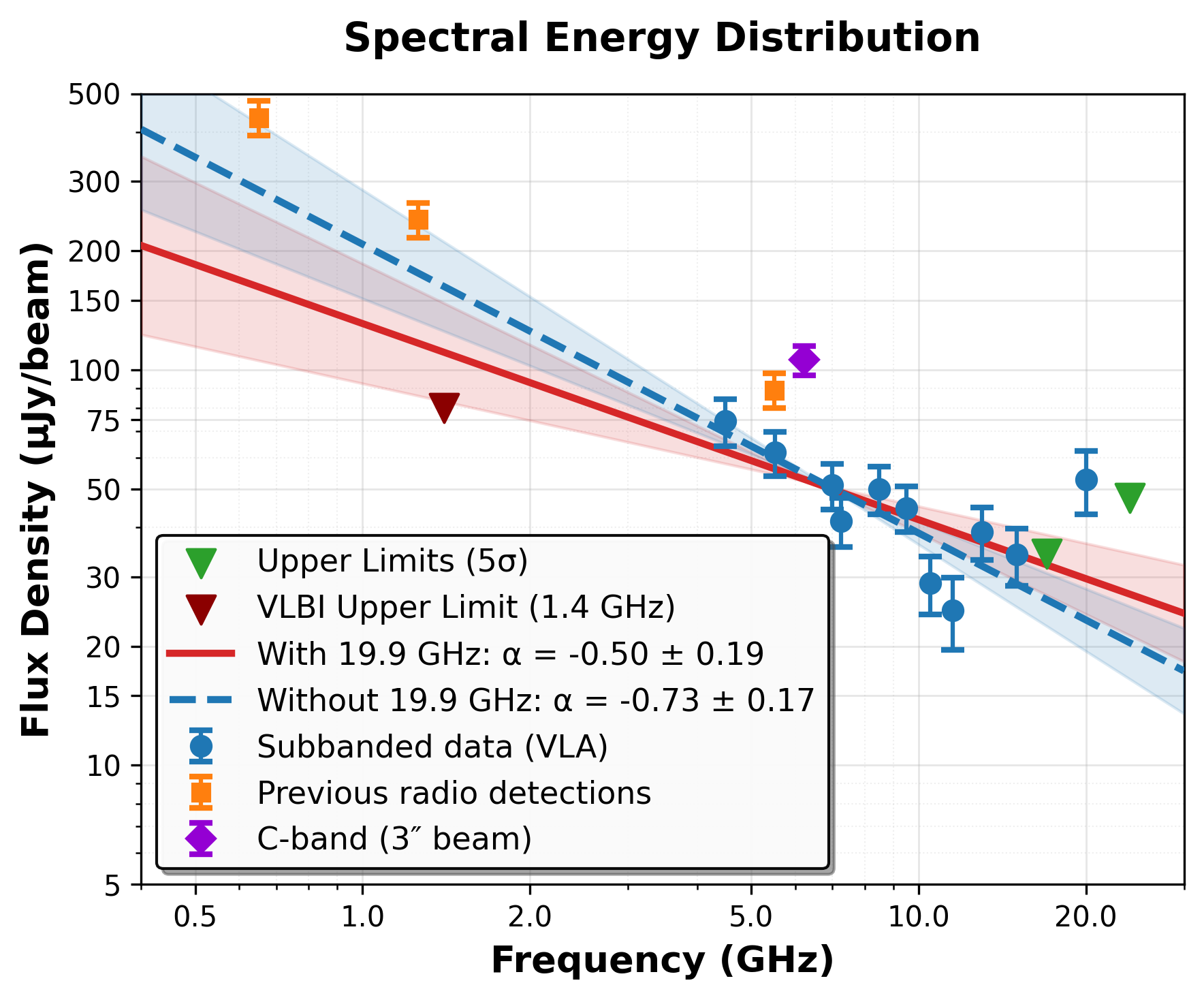}
    \caption{Spectral energy distribution (SED) of the continuum radio counterpart associated with \src{}. Blue circles denote the sub-banded flux densities from our VLA observations, while orange squares represent archival detections from \citet{bhusare_prs} and \citet{2024A&A...690A.219P}. The dark red triangle indicates the $5\sigma$ VLBI upper limit at 1.4 GHz, demonstrating that the source is resolved out on milliarcsecond scales \citep{hewitt2023milliarcsecondlocalisationhyperactiverepeating}. The purple diamond shows the C-band flux density when convolved to a $3''$ synthesized beam to match lower-resolution archival images. Red solid and blue dashed lines represent power-law spectral fits ($S_{\nu} \propto \nu^{\alpha}$) including and excluding the 19.9 GHz outlier, respectively; shaded regions indicate $1\sigma$ uncertainties. The steep spectral index ($\alpha \approx -0.73$) derived by excluding the calibration artifact at 19.9 GHz is consistent with non-thermal emission from a star-forming region.}
    \label{fig:spectra}
\end{figure}

\begin{figure}[ht!]
    \centering
    \includegraphics[width=0.49\textwidth]{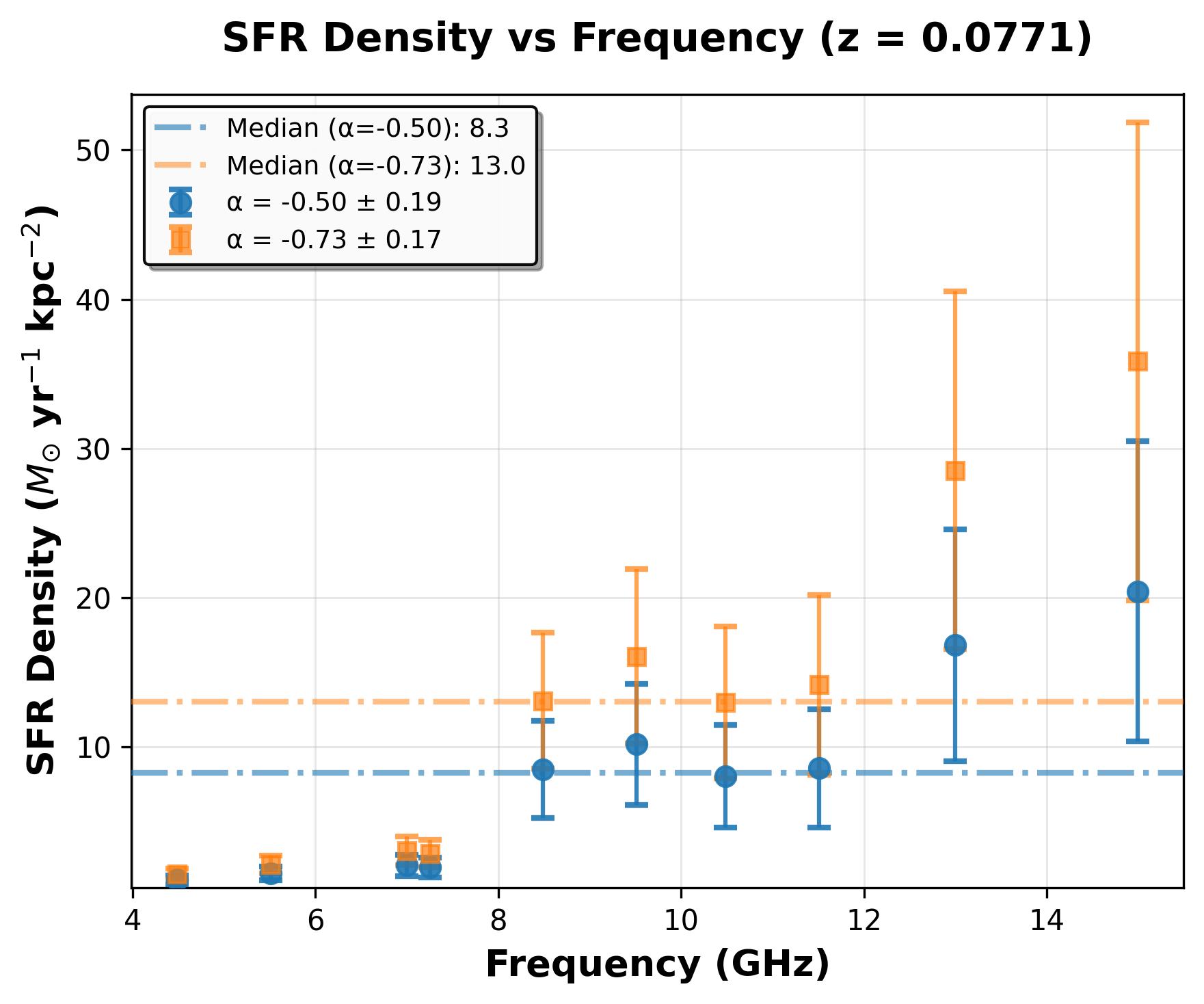}
    \caption{Lower limit on $\Sigma_{\text{SFR}}$ as a function of the observing frequency. The increase in $\Sigma_{\text{SFR}}$ at higher frequencies is a direct consequence of the smaller synthesized beam sizes, which provide better upper limits on the physical area of the unresolved source (for the absolute SFR measurements, which do not scale with beam size, see Figure~\ref{fig:sfr_comparison} in the Appendix \ref{appendix:sfr_stability}). While the measured flux density at 19.9~GHz carries some ambiguity due to a known calibration artifact (see Appendix \ref{appendix:calibration_artifact} and Section~5.2), the geometric constraints provided by the synthesized beam at this frequency remain robust. Blue circles represent $\alpha =- 0.50 \pm 0.19$, derived including the 19.9~GHz anomaly and orange squares show $\alpha = -0.73 \pm 0.17$, excluding the 19.9~GHz anomaly. The shaded regions represent $1\sigma$ uncertainties.}
   
    \label{fig:sfr_density_vs_frequency_zoomed_publication.}
\end{figure}

\section{Results} 
\subsection{Spatial Association with \src{}}

As detailed in Table \ref{tab:astrometry}, we find that the differences in right ascension ($\Delta\text{RA}$) and declination ($\Delta\text{Dec}$) of the continuum source and the FRB are significantly smaller than the synthesized beam's minor axis across all bands. The largest deviation occurs in the C-band, where $\Delta\text{RA} \approx 164$ mas, yet this remains well within the $270$ mas minor beam width (noting that the C-band beam was notably elongated). Our most precise agreement was found in the K-band (~22 GHz), with offsets of only $4.76$ mas and $0.62$ mas, compared to a minor beam size of $90$ mas. These results confirm that the continuum radio emission at all frequencies is spatially coincident with the FRB position. Furthermore, FRB location is approximately 300 mas away from the host galaxy's center.

To assess the chance coincidence probability ($P_{\text{cc}}$) of the FRB position randomly aligning with the detected radio source, we estimate the expected background source density. \citet{2024ApJ...972...89J} conducted a deep $10$~GHz survey of the Great Observatories Origins Deep Survey-North (GOODS-N) field using the VLA, reaching rms noise level of $671$~nJy~beam$^{-1}$. They report a source count of $256$ over an area of $297$~arcmin$^{2}$, which yields a surface number density of $\approx 2.39 \times 10^{-4}$~sources~arcsec$^{-2}$. To establish a highly conservative estimate, we adopt a coincidence radius of $340$~mas, corresponding to twice the synthesized beam size in the X-band. Under the assumption of a Poisson spatial distribution, we calculate a chance coincidence probability of only $P_{\text{cc}} \approx 0.009\%$. This exceedingly low probability robustly disfavors a random alignment.

This alignment confirms that the detected emission is spatially co-located with \src{} and is distinct from the general star-forming activity at the galactic nucleus and thus, disfavors an Active Galactic Nucleus (AGN) origin. 

\subsection{The Spectral Nature of the Radio Counterpart}

To distinguish between a compact PRS powered by a central engine and a localised star forming region, we analyse the spectral energy distribution (SED) and the spatial constraints derived from the VLA-VLBI comparison. The continuum images at different frequencies are shown in Figure \ref{fig:Image}. The corresponding flux densities are mentioned in Table \ref{tab:flux_measurements}, and the resulting SED is shown in Figure \ref{fig:spectra}. We also report non-detections in the 16.9 GHz and 23.9 GHz sub-bands, where the source flux density is likely below the sensitivity limit of our observations. A power-law fit ($S_{\nu} \propto \nu^{\alpha}$) to all data points yields a spectral index of $\alpha = -0.50 \pm 0.19$. 

As mentioned earlier, the measurement at 19.9 GHz had calibration issue (see Section \ref{sec:flux_density_of_source} and the Appendix \ref{appendix:calibration_artifact}). Fitting the flux density without 19.9 GHz yields a steeper spectral index of $\alpha = -0.73 \pm 0.17$. This value is consistent with non-thermal radiation from star-forming regions. We also attempted to correct the artifact by calculating the ratio of the observed spike to the expected modeled flux of the phase calibrator at 19.9 GHz. Applying this scaling factor ($\approx 0.69 \times S_{\text{measured}}$) to the target flux density at the same frequency and re-fitting the SED yields a spectral index of $\alpha = -0.65 \pm 0.15$. This corrected index remains consistent with $\alpha = -0.73$. Notably, the $5\sigma$ upper limit at 16.9 GHz lies within the $1\sigma$ error bar of the predicted flux, while the 23.9 GHz upper limit is significantly displaced above the extrapolated curve. 

\subsection{Effect of refractive interstellar scintillation}

To assess whether Galactic refractive interstellar scintillation (RISS) could significantly perturb the observed SED, we estimated the expected refractive modulation along the line of sight to FRB~20220912A using the NE2001 Galactic electron density model \citep{2002astro.ph..7156C}. For the source coordinates $(l,b)=(106.07^\circ,-10.78^\circ)$, the model predicts a transition frequency of $\nu_t = 20.2$~GHz, placing our 4--20~GHz VLA observations in the strong-scattering regime. In this regime, the modulation index for an extended source is suppressed relative to the point-source value according to
\begin{equation}
m_{\rm eff} = m_R \left(\frac{\theta_r}{\theta_s}\right)^{7/6},
\end{equation}
where $m_R$ is the refractive modulation index for a point source, $\theta_r$ is the refractive angular scale, and $\theta_s$ is the intrinsic angular size of the source \citep{1990ARA&A..28..561R,1998MNRAS.294..307W}. Using the NE2001-derived scattering parameters, we obtain point-source refractive modulation indices ranging from $m_R \sim 0.4$ at 4.5~GHz to $m_R \sim 0.8$ at 15~GHz, with corresponding refractive angular scales of only $\theta_r \sim 2$--40~$\mu$as across our observing bands. However, the VLBI non-detection reported by \citet{hewitt2023milliarcsecondlocalisationhyperactiverepeating}, even after tapering to a $50$~mas beam, implies that the radio emission is resolved out on angular scales $\gtrsim 50$~mas, corresponding to a physical diameter $\gtrsim 75$~pc at $z=0.0771$. This size constraint exceeds the refractive angular scale by approximately three to four orders of magnitude ($\theta_s/\theta_r \sim 10^3$--$10^4$). Substituting these values into the finite-source suppression relation yields effective modulation amplitudes of only $m_{\rm eff} \sim 10^{-5}$--$10^{-4}$ across the observed frequency range. Therefore, Galactic RISS is expected to contribute less than $\sim0.01$ variability to the measured flux densities and cannot meaningfully affect the observed SED or bias the inferred spectral index of $\alpha \approx -0.73$. This confirms that the steep spectrum is intrinsic to the source.


\subsection{Star Formation Rate and Surface Density}

We calculated the Star Formation Rate (SFR) for each detected sub-band using the radio-luminosity-to-SFR relation described in \cite{10.1111/j.1365-2966.2009.15073.x}.

\begin{multline}
\text{SFR} \left[ M_{\odot} \,\text{yr}^{-1} \right] = 
\frac{0.066}{1+z} \times D_{L, \text{Mpc}}^2 \\
\times \left( \frac{(1+z)\,\nu_{\text{obs}}}{1.4} \right)^{-\alpha}
\times S_{\nu, \text{Jy}}
\end{multline}

In this equation, $z$ represents the redshift of $0.0771$, $D_{L, \text{Mpc}}$ is the luminosity distance in Mpc, $\nu_{\text{obs}}$ is the observed frequency in GHz, $S_{\nu, \text{Jy}}$ is the observed peak flux density in Jy, and $\alpha$ is the spectral index\footnote{The negative sign before the spectral index $\alpha$ in the SFR equation is introduced to remain consistent with our convention where a steep spectrum is represented by a negative value.}.

To calculate luminosity distance, we adopted the Planck 2018 cosmological parameters \citep{plank18}. Using \texttt{astropy}, the host galaxy redshift of $z = 0.0771$ \citep{Ravi_2023} yields a luminosity distance of $D_L \approx 361.25$ Mpc.

We calculated the weighted mean SFR using spectral indices derived both with and without the inclusion of the 19.9 GHz flux density. Utilizing the spectral index obtained by including the 19.9 GHz point ($\alpha = -0.50 \pm 0.19$), we find a weighted mean SFR of $0.9 \pm 0.1$ $M_{\odot}$ yr$^{-1}$. When the 19.9 GHz point is excluded from the spectral index fit ($\alpha = -0.73 \pm 0.17$), the weighted mean SFR is $1.3 \pm 0.1$ $M_{\odot}$ yr$^{-1}$. These values, along with their associated uncertainties, are summarized in Figure \ref{fig:sfr_comparison} in the appendix \ref{appendix:sfr_stability}.

The star-formation rate surface density, $\Sigma_{\text{SFR}}$, is defined as the star-formation rate ($M_{\odot}\,\text{yr}^{-1}$) normalized by the physical area ($kpc^2$) of the emitting region. We characterize the source by calculating the star formation rate surface density ($\Sigma_{\text{SFR}}$) for each sub-band. Since the radio source remains unresolved across all frequencies, the synthesized beam area in each band defines a strict upper limit on the physical extent of the emitting region. Consequently, the derived $\Sigma_{\text{SFR}}$ values represent conservative lower limits on the actual star formation density. Our beam areas range from $\approx 1.01$~kpc$^2$ at 4.5~GHz down to $\approx 0.028$~kpc$^2$ at 20~GHz. The $\Sigma_{\text{SFR}} $ is $ \gtrsim 1.1 \pm 0.3~M_{\odot}~\text{yr}^{-1}~\text{kpc}^{-2}$ at 4.48 GHz and $ \gtrsim 20.4 \pm 10.7~M_{\odot}~\text{yr}^{-1}~\text{kpc}^{-2}$ at 14.9 GHz. For a spectral index of $\alpha = -0.73 \pm 0.17$ (which excludes the 19.9~GHz artifact), the derived median value of $\Sigma_{\text{SFR}}$ is $13.0~M_{\odot}~\text{yr}^{-1}~\text{kpc}^{-2}$. As shown in Figure~\ref{fig:sfr_density_vs_frequency_zoomed_publication.}, we observe an apparent increase in $\Sigma_{\text{SFR}}$ with frequency. This trend is primarily driven by improved spatial constraints with frequency rather than intrinsic spectral evolution. Even when considering the flatter spectral index ($\alpha = -0.50$), the median $\Sigma_{\text{SFR}}$ remains high at $8.3~M_{\odot}~\text{yr}^{-1}~\text{kpc}^{-2}$. For the subsequent discussion, we adopt the median as the characteristic minimum surface density, i.e., $\Sigma_{\text{SFR}} \gtrsim 13~M_{\odot}~\text{yr}^{-1}~\text{kpc}^{-2}$, for the FRB~20220912A environment.

\section{Discussion}

\subsection{Constraints on the Emission Scale} 
The spatial coincidence of a radio source with \src{} allows for two primary interpretations: 
\begin{itemize}
    \item A Compact ( $d < 25$ pc ) PRS: A central-engine-powered source (e.g., a pulsar/magnetar wind nebula or supernova remnant) which generally exhibits a flat spectrum. However, an inverted spectrum ($\alpha > 0$) at high frequencies has also been observed, at least for one FRB \citep[for FRB 20201124A;][]{bruni2024nebularoriginpersistentradio}. 
    \item A Star-Forming Region: A localized knot of star formation within the host galaxy, characterized by a steep non-thermal spectrum ($\alpha \approx -0.7$). 
\end{itemize}

By comparing the VLA A-configuration detections (this work) with the lack of a counterpart in high-resolution VLBI observations \citep{hewitt2023milliarcsecondlocalisationhyperactiverepeating}, we can constrain the source size. The evidence for an extended source comes from the ``missing" flux at L-band. Extrapolating our VLA peak flux density measurements to 1.4 GHz, even using the flatter of our two $\alpha$ estimates (i.e., $\alpha = -0.50$), predicts a flux density of $111 \pm 18~\mu\text{Jy}$ (shaded region, Figure \ref{fig:spectra}). This is more than the $5\sigma$ VLBI upper limit of $80~\mu\text{Jy}$ \citep{hewitt2023milliarcsecondlocalisationhyperactiverepeating}. To investigate whether this non-detection was simply a result of the emission being resolved by the $\sim 30$~mas VLBI beam, we convolved the VLBI data to a coarser $50$~mas resolution. Even with this larger synthetic beam, no emission was recovered ($5\sigma < 30~\mu\text{Jy}$; see Figure \ref{fig:pos_match}). Given the $6~\mu\text{Jy}$ RMS noise, a compact $111~\mu\text{Jy}$ source should have been detected with high significance ($\text{SNR} > 15$). The absence of a detection, even at 50~mas, suggests the emission to be distributed over scales larger than the maximum angular extent VLBI array can probe. At $z = 0.077$, this 50~mas limit implies a physical diameter $d \geq 75$~pc.

At the other end of the scale, we can compare our high-resolution A-configuration data with lower-resolution archival images. When convolved to a $3''$ beam to match D-configuration VLA archival data, our C-band image yields a peak flux density of $106 \pm 9~\mu\text{Jy}$, consistent with the $89 \pm 9~\mu\text{Jy}$ reported in \cite{bhusare_prs} within $1.3 \sigma$. The peak flux density with $3''$ beam is significantly higher than the peak flux density at high resolution, indicating the presence of extended emission from the host galaxy, in addition to the compact knot at the FRB location. One potential concern is whether the elongated beam in our C-band observations ($0.65''\times 0.26''$) allows for contamination from the host galaxy's nuclear emission. The FRB position is offset from the host galaxy center by $\approx 300$~mas ($\approx 450$~pc). While our C-band image (full band, using robust=0.5 weighting) has a major axis of $0.65''$, we also produced images using uniform weighting. This improved the C-band resolution to $0.40'' \times 0.15''$. At this resolution, there is less spatial overlap between the synthesized beam at the FRB position and the galactic center. Crucially, the peak flux density at the FRB position remains consistent across different weighting schemes. Furthermore, even if we recalculate the spectral index using only the X-band and Ku-band data, where the synthesized beams are smaller than 300~mas, we still derive a steep spectral index of $\alpha \approx -0.7$. This confirms that the steep spectrum is an intrinsic property of \src{} environment and not a result of host galaxy's nuclear contamination.

While the 19.9 GHz flux density is affected by a calibration artifact, the detection itself remains spatially coincident with the FRB and can be used to set a conservative upper limit on the source size; at this frequency, the $0.13'' \times 0.12''$ synthesized beam constrains the emitting region to a physical scale of $d < 190 \times 178$ pc. Combined with the lower limit from the VLBI resolved-out analysis ($d > 75$ pc), this limits the physical extent of the source to approximately $75$–$190$ pc, a scale consistent with a massive star forming region \citep{Kennicutt_Jr_1998, Heiderman_et_al}. Note that the physical size constraints derived here utilize the beam dimensions from our naturally weighted sub-band images (Table \ref{tab:beam_properties}), which are intrinsically larger than the full-bandwidth beams (Table \ref{tab:astrometry}) used for astrometry.

\subsection{Brightness Temperature and Emission Mechanism}

The brightness temperature ($T_b$) serves as a critical diagnostic to distinguish between a thermal/non-thermal star-forming region and a coherent emission. We calculated $T_b$ for all sub-bands using the Rayleigh-Jeans approximation:
$$T_b \approx 1.22 \times 10^3 \frac{S_{\nu}}{\nu^2 \theta_{maj} \theta_{min}} \text{ K}$$

Here, $S_{\nu}$ is the peak flux density and $\theta_{\text{maj}}$ and $\theta_{\text{min}}$ are the full width at half-maximum (FWHM) of the synthesized major and minor axes of the beam, respectively. 
Across our 4--24 GHz VLA observations, we consistently find $T_b < 100$ K. Since the source remains unresolved, the synthesized beam (in the individual sub-bands) is being used to derive conservative Tb constraints. Even if we assume the source to be as small as the lower limit of $\sim 75$ pc imposed by the VLBI analysis, the brightness temperature at 1.4 GHz would only reach $T_b \approx 10^4$ K. This is significantly lower than $T_b \gtrsim 10^7$ K observed in the PRSs of FRB 20121102A and FRB 20190520B \citep{Chatterjee_2017,Niu_2022}. High $T_b$ values in those sources indicate highly compact, dense, and potentially magnetized plasma environments. Our low $T_b$, combined with the steep spectral index ($\alpha \approx -0.7$), points to an optically thin supernova remnant or HII region.

\begin{table*}[ht!]
\centering
\caption{Comparison of Radio Environments of FRB 20220912A and Known PRS-Associated FRBs}
\label{tab:prs_comparison}
\renewcommand{\arraystretch}{1.3}
\small 
\begin{tabular}{llccccc} 
\hline
\hline
\textbf{FRB Name} & \textbf{Emission Type} & \textbf{Size (Compact)} & \textbf{Size (Extended)} & \textbf{$\alpha$} & \textbf{RM} & \textbf{$\Sigma_{\text{SFR}}$} \\
 & & & & & ($\text{rad m}^{-2}$) & ($M_{\odot}\text{ yr}^{-1}\text{ kpc}^{-2}$) \\
\hline
20220912A$^{a}$ & Extended SFR & [--] & $75$--$190$ pc & $-0.73$ & $\sim 0$ & $\gtrsim 13$ \\
20121102A$^{b}$ & Compact PRS & $< 1$ pc & $2 \times 0.68$ kpc & $-0.15$ & $\sim 10^5$ & $\gtrsim 0.17$ \\
20190520B$^{c}$ & Compact PRS & $\lesssim 23$ pc & [--] & $-0.33 $ & $\sim 10^4$  & [--] \\
20201124A$^{d}$ & Ext./Unconfirmed & $< 700$ pc & [--] & $ 1$ & $\sim 10^3$ & $\gtrsim 1.1$ \\
20240114A$^{e}$ & Compact PRS & $< 4$ pc & [--] & $-0.14$ & $\sim 10^2$ & [--] \\
20190417A$^{f}$ & Compact PRS & $< 23$ pc & [--] & $-0.19$ & $\sim 5000$ & [--] \\
\hline
\multicolumn{7}{l}{\textbf{Note.} $\alpha$ represents the radio spectral index ($S_{\nu} \propto \nu^{\alpha}$). The `Size (Compact)' column denotes} \\
\multicolumn{7}{l}{upper limits on the physical size of the central engine-powered PRS, while `Size (Extended)' represents the physical scale} \\
\multicolumn{7}{l}{of the associated radio source.} \\
\multicolumn{7}{l}{\textbf{References:} (a) This Work; (b) \citet{2017ApJ...834L...8M}, \citet{Chatterjee_2017}, \citet{2025arXiv250623861B}}\\
\multicolumn{7}{l}{ (c) \citet{Niu_2022} , \citet{2023ApJ...958L..19B}, \citet{2023ApJ...959...89Z}}\\
\multicolumn{7}{l}{(d) \citet{bruni2024nebularoriginpersistentradio}, (e) \citet{bhusare_prs, bruni2024discoveryprsassociatedfrb} }\\
\multicolumn{7}{l}{(f) \citet{2026ApJ...996L..16M,2026arXiv260403429B}.} \\
\hline
\end{tabular}
\end{table*}

\subsection{Local Star Formation and Implications for the Progenitor}
The derived lower limit on the SFR surface density ($\Sigma_{\text{SFR}} \gtrsim 13~M_{\odot}~\text{yr}^{-1}~\text{kpc}^{-2}$) provides a high-resolution look at the local environment that previous studies could not access. Initial host galaxy studies by \cite{Ravi_2023} estimated a global SFR $\gtrsim 0.1~M_{\odot}~\text{yr}^{-1}$. Our results show that the star formation is not uniformly distributed but is instead concentrated in a compact knot coincident with the FRB position.

Our derived median value of $\Sigma_{\text{SFR}} $ places the FRB environment several orders of magnitude above the diffuse emission of typical spiral disks, which generally range from $0$ to $0.1~M_{\odot}~\text{yr}^{-1}~\text{kpc}^{-2}$ \citep{Kennicutt_Jr_1998}. Instead, the environment of \src{} is more consistent with the ``circumnuclear regions" of starburst galaxies, which typically exhibit densities between $1$ and $1000~M_{\odot}~\text{yr}^{-1}~\text{kpc}^{-2}$ \citep{Kennicutt_Jr_1998}. Furthermore, the surface density of this star-forming knot is comparable with active star forming regions. For example, the 30~Doradus region in the Large Magellanic Cloud, one of the most active star forming complexes in the Local Group, has an estimated $\Sigma_{\text{SFR}} \approx 3.2 \pm 1.7~M_{\odot}~\text{yr}^{-1}~\text{kpc}^{-2}$ (derived from \citealt{Doran_et_al}). Our median value is also higher than the average density of Young Stellar Object (YSO) clusters in local molecular clouds ($7.7 \pm 2.0~M_{\odot}~\text{yr}^{-1}~\text{kpc}^{-2}$), though it sits comfortably within the observed peak range for such clusters, which can reach $\sim 79.6~M_{\odot}~\text{yr}^{-1}~\text{kpc}^{-2}$ \citep{Heiderman_et_al}.

Many FRBs are found in regions where the local star formation rate is comparable to the average SFR of their host galaxies. Notably, FRB~20121102A and FRB~20201124A are both located within star-forming regions and are associated with PRS. FRB~20121102A resides in a region with a SFR surface density of \(\Sigma_{\text{SFR}} \approx 0.17~M_{\odot}~\text{yr}^{-1}~\text{kpc}^{-2}\), while FRB~20201124A is located in a region with 
\(\Sigma_{\text{SFR}} \approx 1.1~M_{\odot}~\text{yr}^{-1}~\text{kpc}^{-2}\), 
among the highest values measured for known FRBs \citep{2017ApJ...843L...8B, 2021A&A...656L..15P, Dong24,bruni2024nebularoriginpersistentradio}. 

In contrast, FRB~20190608B, is situated in a region with much lower star formation activity, 
\(\Sigma_{\text{SFR}} \approx 1.5 \times 10^{-2}~M_{\odot}~\text{yr}^{-1}~\text{kpc}^{-2}\) 
\citep{2021ApJ...922..173C}. Similarly, FRB~20180916B is offset by \(\sim 250\) pc from the nearest star-forming region \citep{2021ApJ...908L..12T}. The nearby non-repeater FRB~20250316A also lies close to a star-forming region, but is displaced by \(190 \pm 20\) pc \citep{2025ApJ...989L..48C}. Comparing with some of the FRBs with known local star-formation densities (see Table~\ref{tab:prs_comparison}), \src{} resides in the highest SFR surface density. Table \ref{tab:prs_comparison} also provides a summary of the other radio properties of the source associated with \src{}, along with a comparison to FRBs that are known to have associated PRSs.

By placing \src{} in such a high-density knot ($\Sigma_{\text{SFR}} \gtrsim 13~M_{\odot}~\text{yr}^{-1}~\text{kpc}^{-2}$), we reinforce the ``young magnetar" progenitor theory, as these environments are the primary birthplaces of the massive stars ($>8~M_{\odot}$) that end in core-collapse supernovae. The absence of a compact PRS and the low observed RM suggest that the immediate circumburst environment is less dense or more evolved than the extreme environments of FRB~20121102A or FRB~20190520B.

While our radio data provide a self-consistent argument for a compact star-forming region, more direct probes of the star formation rate, kinematics of the region, and metallicity require multi-wavelength follow-up. Specifically, spatially resolved mapping of the $H\alpha$ and $[OIII]$ emission at sub-arcsecond scales, using Integral Field Unit (IFU) observations, would be extremely useful.

\section{Summary}

We have conducted a broadband (4--26~GHz) study of the local environment of the hyperactive repeating \src{} using the VLA in its most extended A and BnA configurations. Our primary findings are as follows.

\begin{itemize}
    \item \textbf{Detection of a Compact Radio Counterpart:} We have detected radio continuum emission spatially coincident with the FRB position across all the observed receiver bands, spanning $4$--$26$~GHz range. The high-frequency observations constrain the unresolved source to an angular size of $\lesssim 130 \times 120$ mas. The source exhibits a peak flux density of $44.8 \pm 3.9~\mu\text{Jy~beam}^{-1}$ at $9.51$~GHz with a spectral index of -0.73. The source is distinctly offset by $\approx 300$~mas ($\approx 450$~pc) from the host galaxy's center, confirming the emission to be in a separate region coincident with the FRB, rather than being a part of the central galactic nucleus.
    
    \item \textbf{A Star-Forming Region:} By combining our VLA spatial constraints with the non-detection from archival high-resolution VLBI observations \citep{hewitt2023milliarcsecondlocalisationhyperactiverepeating}, we limit the physical diameter of the emitting region between 75~pc and 190~pc. This spatial scale, combined with a steep non-thermal spectral index ($\alpha \approx -0.73$) and a low brightness temperature ($T_b < 100$~K), strongly favors the source as a compact star-forming knot.
    
    \item \textbf{Intense Local Star Formation:} We derive a lower limit on localized star-formation rate surface density of $\Sigma_{\text{SFR}} \gtrsim 13~M_{\odot}~\text{yr}^{-1}~\text{kpc}^{-2}$. This is exceptionally high, placing the immediate environment of \src{} on par with the most intense star-forming complexes in the local universe, such as the 30~Doradus region \citep{Doran_et_al}.
\end{itemize}

Our analysis is highly consistent with \src{} residing in a high-density stellar nursery characterized by a high local star formation rate surface density. Such environments are the birthplaces of massive stars and therefore naturally favor progenitor scenarios involving young magnetars formed through core-collapse supernovae. Our results also highlight the importance of combining broadband radio observation with high-resolution imaging to distinguish between central engine powered PRS and radio emission arising from the local star-forming environment.

\begin{acknowledgments}
YB and YM would like to thank Nissim Kanekar for the insightful discussions on the Imaging analysis. YB sincerely acknowledges Arvind Balasubramanian for helping with the VLA observations and Benito Marcote for valuable insights regarding the understanding of the VLBI dataset. YM acknowledges support from the Department of Science and Technology via the Science and Engineering Research Board Startup Research Grant (SRG/2023/002657). Y.D. is supported by the National Science Foundation Graduate Research Fellowship under grant No. DGE-2234667. We would like to thank the Centre Director and the observatory for the prompt time allocation and scheduling of our observations. The National Radio Astronomy Observatory and Green Bank Observatory are facilities of the U.S. National Science Foundation operated under cooperative agreement by Associated Universities, Inc. This research has made use of the NASA/IPAC Extragalactic Database (NED), which is operated by the Jet Propulsion Laboratory, California Institute of Technology, under contract with the National Aeronautics and Space Administration.
\end{acknowledgments}

\bibliography{ref}{}
\bibliographystyle{aasjournalv7}

\appendix 
\section{The 19.9 GHz Flux Density Calibration Artifact} \label{appendix:calibration_artifact}

\begin{figure}[ht!]
    \centering
    \includegraphics[width=0.7\textwidth]{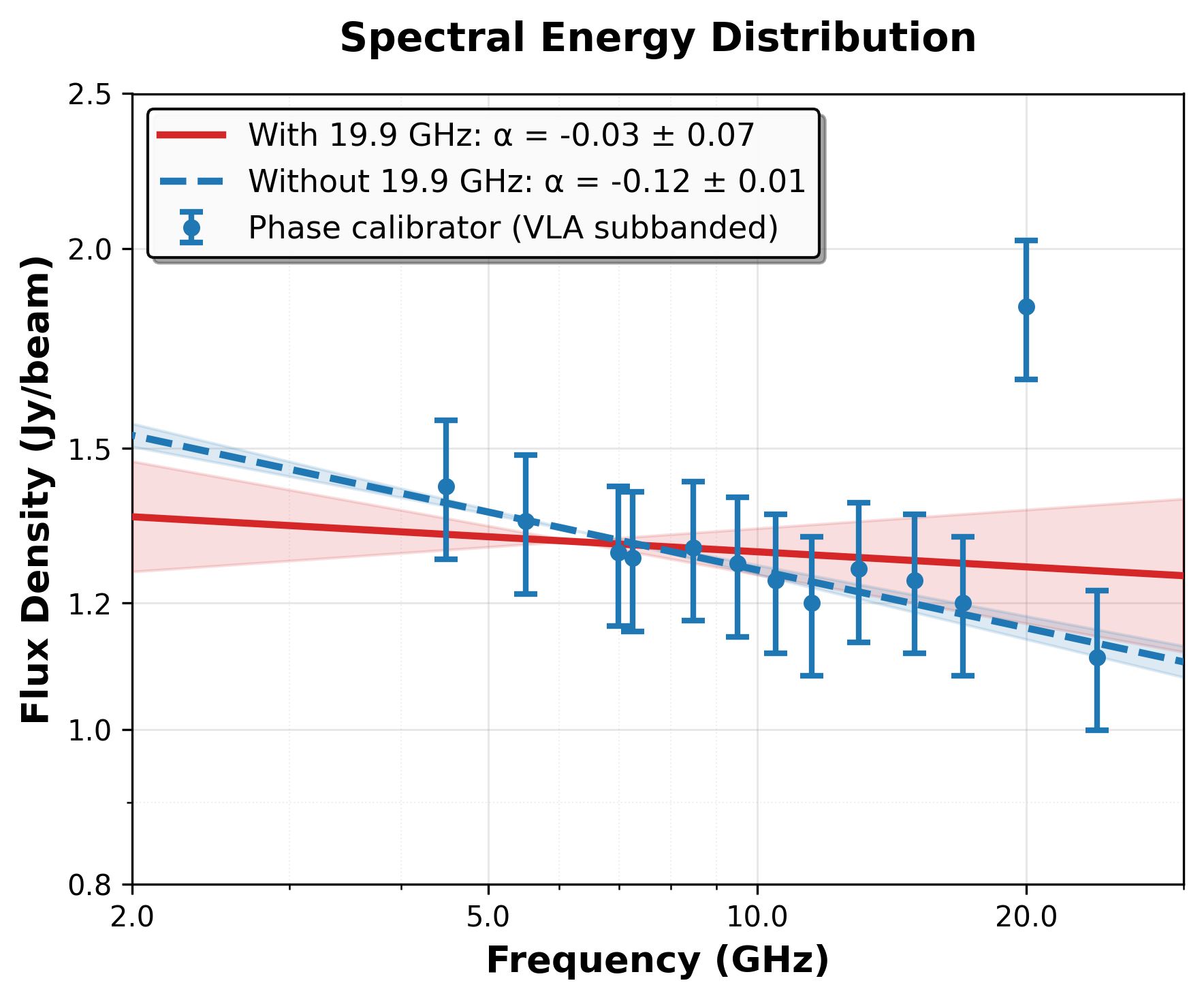}
    \caption{Spectral energy distribution (SED) of the phase calibrator (J2322+509). Blue points represent the measured flux density in each VLA sub-band. Similar to the radio source (see Figure~\ref{fig:spectra}), a significant flux density enhancement is observed at $\sim$19.99 GHz. This correlated feature in a known stable calibrator confirms that the high-frequency spike is a systematic calibration artifact, likely due to water line absorption at 22.2 GHz rather than an intrinsic property of the FRB 20220912A environment. The blue dashed line indicates the preferred spectral fit ($\alpha = -0.121 \pm 0.014$) excluding this outlier, consistent with the expected flat-spectrum nature of the calibrator.}
    \label{fig:spectra_phase_cal}
\end{figure}

We observe a significant flux density enhancement at 19.9 GHz. To determine whether this feature is intrinsic to the source or a systematic calibration error, we analyzed the spectral behavior of the phase calibrator (J2322+509) observed during the same session. As shown in Figure \ref{fig:spectra_phase_cal}, the phase calibrator exhibits a flux enhancement at the exact same frequency. This artificial spike may arises from atmospheric opacity variations near the 22.2 GHz water vapor absorption line. As the primary flux density calibrator (3C48) and the target field are separated by an angular distance of $\Delta\theta \approx 30.35^\circ$ on the sky, they are sampled through distinct atmospheric columns with significantly varying precipitable water vapor (PWV) contents. Furthermore, at the observation epoch, the primary flux calibrator (3C48) was at a lower altitude ($\approx 20^\circ$), whereas the phase calibrator (J2322+509) was at a higher elevation ($\approx 48^\circ$). If the flux calibrator experiences stronger water line absorption than the target field, the calibration derives anomalously high gain solutions to compensate for the attenuated signal. When these inflated gain solutions are transferred to the target and phase calibrator, their measured flux densities near the water line would get artificially boosted.

\section{Star Formation Rate Calculation}
\label{appendix:sfr_stability}

\begin{figure*}[htbphtbp]
    \centering
    \includegraphics[width=0.7\textwidth]{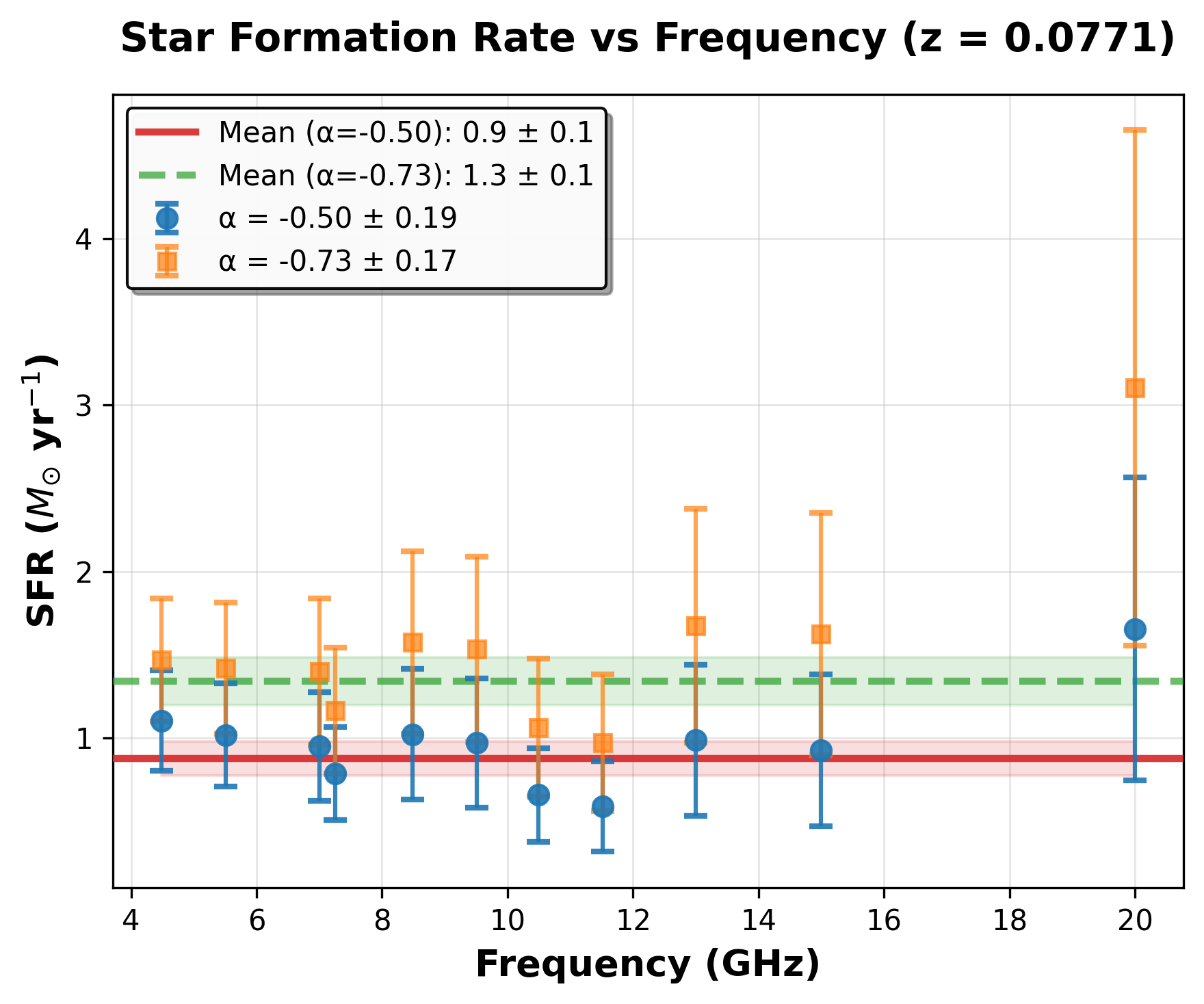}
    \caption{Derived Star Formation Rate (SFR) as a function of observing frequency for the local environment of \src{}. Unlike the SFR surface density ($\Sigma_{\text{SFR}}$) shown in Figure~4, SFR does not strongly depend on frequency-dependent synthesized beam area and is stable across the spectrum by design. Data points are plotted for two derived spectral indices: blue circles ($\alpha = -0.50 \pm 0.19$, derived including the 19.9~GHz anomaly) and orange squares ($\alpha = -0.73 \pm 0.17$, preferred index excluding the 19.9~GHz anomaly). Horizontal lines indicate the corresponding weighted mean SFR values: the solid red line shows the mean based on $\alpha= -0.50$ points ($0.9 \pm 0.1~M_{\odot}~\text{yr}^{-1}$), while the dashed green line shows the preferred mean based on $\alpha= -0.73$ points ($1.3 \pm 0.1~M_{\odot}~\text{yr}^{-1}$); shaded regions represent $1\sigma$ uncertainties.}
    \label{fig:sfr_comparison}
\end{figure*}

\begin{table*}[htbp]
    \centering
    \caption{Synthesized Beam Properties per Sub-band}
    \label{tab:beam_properties}
    \begin{tabular}{lccc}
        \toprule
        \textbf{Frequency} & \textbf{Major Axis} & \textbf{Minor Axis} & \textbf{Position Angle} \\ 
        \textbf{(GHz)} & \textbf{($^{\prime\prime}$)} & \textbf{($^{\prime\prime}$)} & \textbf{($^\circ$)} \\ \midrule
        4.48  & 1.19 & 0.48 & 65.7 \\
        5.51  & 0.99 & 0.38 & 65.8 \\
        6.99  & 0.82 & 0.31 & 62.8 \\
        7.25  & 0.76 & 0.30 & 63.2 \\
        8.48  & 0.28 & 0.24 & 14.0 \\
        9.51  & 0.25 & 0.22 & 13.8 \\
        10.48 & 0.22 & 0.21 & 43.4 \\
        11.51 & 0.20 & 0.19 & 39.6 \\
        12.99 & 0.21 & 0.16 & 50.9 \\
        14.99 & 0.18 & 0.14 & 53.1 \\
        19.99 & 0.13 & 0.12 & -44.6 \\ \bottomrule
    \end{tabular}
    \tablecomments{Beam dimensions corresponding to the detections listed in Table~\ref{tab:flux_measurements}. These dimensions define the upper limits on the physical size of the unresolved continuum source used to calculate the brightness temperature ($T_b$) and star formation rate surface density ($\Sigma_{\text{SFR}}$).}
\end{table*}

To verify that our physical interpretations are not biased, we calculated the Star Formation Rate (SFR) across the entire observed spectrum. Figure~\ref{fig:sfr_comparison} displays the derived SFR for each sub-band. Because a fixed spectral index ($\alpha$) is assumed across the band, the SFR remains flat by construction. Furthermore, the continuum radio counterpart remains unresolved across all the sub-bands, and the spatial footprint of the synthesized beam at each individual frequency configuration sets a strict physical upper limit on the true extent of the emitting region. Table~\ref{tab:beam_properties} details the major axis, minor axis, and position angle for each sub-banded image. 
\end{CJK*}
\end{document}